**Title:** Optimization of acquisition parameters for cortical inhomogeneous magnetization transfer (ihMT) imaging using a rapid gradient echo readout

**Authors:**

Christopher D. Rowley (1 and 2), Jennifer S.W. Campbell (1), Ilana R. Leppert (1), Mark C. Nelson (1 and 2), G. Bruce Pike (3), Christine L. Tardif (1 and 2 and 4)

(1) McConnell Brain Imaging Centre, Montreal Neurological Institute and Hospital, McGill University, Montreal, QC, Canada H3A 2B4

(2) Department of Neurology and Neurosurgery, McGill University, Montreal, QC, Canada H3A 2B4

(3) Hotchkiss Brain Institute and Departments of Radiology and Clinical Neuroscience, Cumming School of Medicine, University of Calgary, Calgary, Canada T2N 4N1

(4) Department of Biomedical Engineering, McGill University, Montreal, QC, Canada H3A 2B4

**Abstract:**

*Purpose*: Imaging biomarkers with increased myelin specificity are needed to better understand the complex progression of neurological disorders. Inhomogeneous magnetization transfer (ihMT) imaging is an emergent technique that has a high degree of specificity for myelin content but suffers from low signal-to-noise ratio (SNR). This study used simulations to determine optimal sequence parameters for ihMT imaging for use in high-resolution cortical mapping.

*Methods:* MT-weighted cortical image intensity and ihMT SNR were simulated using modified Bloch equations for a range of sequence parameters. The acquisition time was limited to 4.5 min/volume. A custom MT-weighted RAGE sequence with center-out k-space encoding was used to enhance SNR at 3 Tesla. Pulsed MT imaging was studied over a range of saturation parameters, and the impact of the turbo-factor on the effective ihMT resolution was investigated. 1 mm isotropic $ihMT_{sat}$ maps were generated in 25 healthy adults.

*Results:* Greater SNR was observed for larger number of bursts consisting of 6-8 saturation pulses each, combined with a high readout turbo-factor. However, that protocol suffered from a point spread function that was more than twice the nominal resolution. For high-resolution cortical imaging, we selected a protocol with a higher effective resolution at the cost of a lower SNR. We present the first group-average $ihMT_{sat}$ whole-brain map at 1 mm isotropic resolution.

*Conclusion:* This study presents the impact of saturation and excitation parameters on $ihMT_{sat}$ SNR and resolution. We demonstrate the feasibility of high-resolution cortical myelin imaging using $ihMT_{sat}$ in less than 20 minutes.

## Introduction:

Magnetization transfer (MT) between hydrogen atoms has long been exploited as a contrast mechanism in MRI (1,2). It arises from the chemical exchange or cross-relaxation between hydrogen in the free water pool and hydrogen attached to macromolecules (1). To generate this contrast, a radiofrequency (RF) pulse is applied at a frequency shifted from the free water resonance frequency to saturate bound hydrogen. Through these exchange processes, magnetization is transferred between the macromolecular pool and the free water pool. The magnetization exchange presents itself as a decrease, or partial saturation, of the free water signal. The extent of this saturation is a function of multiple parameters including the size of the free and bound pools, the exchange rate between the pools, the strength and off-resonance frequency of the MT saturation radiofrequency (RF) pulse, and the relaxation rates of both pools (3).

Inhomogeneous MT (ihMT) contrast arises from the differential saturation between MT-weighted images collected with a single offset frequency (+ or -$\Delta$) and with a dual offset frequency ($\pm\Delta$) saturation scheme that is mirrored around the central resonant frequency (4). This is based on Provotorov theory (5), where different levels of MT saturation are observed for $\pm\Delta$ versus + or -$\Delta$ pulses of equal power, due to the residual dipolar coupling present in some molecules (6). This contrast mechanism was demonstrated to have increased myelin-specificity compared to conventional MT and to increase linearly with a histological stain for myelin content (7,8). As ihMT contrast is weighted by the dipolar relaxation time ($T_{1D}$), increased specificity to myelin content is made possible due to the longer $T_{1D}$ in lipid-rich myelin sheaths compared to cellular proteins (9). Readers are directed to a recent review for a more comprehensive overview of ihMT (10).

MT saturation ($MT_{sat}$) is widely used to quantify the MT effect in a voxel while minimizing the confounding effects of the $T_1$ relaxation and $B_1^+$ field inhomogeneity across the brain (11). The calculation of $MT_{sat}$ requires the measurement of two tissue properties, $T_1$ and $M_0$ (the equilibrium magnetization, a value that is proportional to the proton density), and an MT-weighted image. These values permit the calculation of $MT_{sat}$ as the percent drop in the pseudo-steady state signal due to the MT saturation pulses. This was originally calculated for a case with one excitation pulse per repetition time (TR). If an accelerated acquisition is used with multiple excitation pulses, the pseudo-steady state signal equation must be modified to account for the multiple flip angles applied in each TR (**Figure 1**) (12,13). For high resolution imaging, a multi-line readout (turbo-factor) may be desired to reduce scan time. This is necessary for ihMT imaging as long TRs are required to accommodate the increased SAR demands of the high $B_{1rms}$ pulses that improve the ihMT effect (14). The $B_{1rms}$ for a pulse ($B_{1rms,pulse}$) can be calculated as:

$$B_{1rms,pulse} = B_{1,peak} \cdot \sqrt{\frac{1}{t_{pulse}} \int_{0}^{t_{pulse}} A(t)^2 dt} \tag{1}$$

where $B_{1,peak}$ is the amplitude of the pulse, $t_{pulse}$ is the duration of the pulse, and $A(t)$ is the normalized pulse shape. This can be extended to look at the $B_{1rms}$ over the saturation block ($B_{1rms,sat}$) using (15):

$$B_{1rms,sat} = \sqrt{B_{1rms,pulse}{}^2 \cdot \left(\frac{N_p \cdot t_{pulse}}{t_{sat}}\right)} \tag{2}$$

where $N_p$ is the number of pulses applied over saturation time block $t_{sat}$. The $B_{1rms}$ can be equivalently calculated over the excitation block ($B_{1rms,exc}$), using time $t_{read}$ and the turbo-factor. These values can be used to calculate the $B_{1rms}$ over the TR ($B_{1rms,TR}$) as:

$$B_{1rms,TR} = \sqrt{B_{1rms,sat}{}^2 \cdot \left(\frac{t_{sat}}{TR}\right) + B_{1rms,exc}{}^2 \cdot \left(\frac{t_{read}}{TR}\right)} \tag{3}$$

Combined with our recent $B_1^+$ correction method (16), this work aims to optimize $ihMT_{sat}$ imaging for the cortical grey matter (GM) with a scan time <20 mins. This study uses a rapidly acquired gradient echo (RAGE) readout, as has been previously used for cortical ihMT imaging (12). As ihMT was shown to increase linearly with myelin content, this optimization focused on maximizing ihMT SNR required for high-resolution imaging of the thin and convoluted cortical ribbon. The present study investigates the degree to which bursts of saturation pulses, which facilitate low duty-cycle acquisitions with maximized saturation power, could increase the SNR in high-resolution cortical ihMT imaging. Furthermore, within the constraint of a fixed scan time, we investigate how changes in the turbo-factor of the RAGE readout, required to accommodate different saturation schemes, impacts the point-spread function (PSF).

**Theory:**

Magnetization Transfer Saturation

Here we utilize a two-pool model with dipolar order, which consists of the transverse magnetizations of the water pool ($M_X^A$, $M_Y^A$), the longitudinal magnetizations of the water pool ($M_Z^A$) and bound pool ($M_Z^B$), and the dipolar order of the bound pool ($\beta'$) (illustrated in **Supplementary Material Figure S1**). Modified Bloch equations that incorporate dipolar mediated spin diffusion and chemical exchange can be numerically evaluated for a time varying off-resonance pulse (17–20):

$$\frac{dM_X^A}{dt} = -R_{2A}M_X^A - 2\pi\Delta M_Y^A \tag{4}$$

$$\frac{dM_Y^A}{dt} = -R_{2A}M_Y^A + 2\pi\Delta M_X^A - \omega_1 M_Z^A \tag{5}$$

$$\frac{dM_Z^A}{dt} = R_{1A}(M_0^A - M_Z^A) - kM_0^B M_Z^A + kM_0^A M_Z^B + \omega_1 M_Y^A \tag{6}$$

$$\frac{dM_Z^B}{dt} = R_{1B}(M_0^B - M_Z^B) + kM_0^B M_Z^A - kM_0^A M_Z^B - R_{rfB}(M_Z^B - \Omega\beta') \tag{7}$$

$$\frac{d\beta'}{dt} = \Omega R_{rfB}(M_Z^B - \Omega\beta') - \frac{\beta'}{T_{1D}} \tag{8}$$

The evaluation of this set of coupled equations requires the knowledge of the following variables: the equilibrium longitudinal magnetization of the free water pool ($M_0^A$) and bound pool ($M_0^B$), the longitudinal magnetization relaxation rates $R_{1A}$ ($1/T_{1A}$), $R_{1B}$ ($1/T_{1B}$), the dipolar order decay time ($T_{1D}$), the transverse magnetization relaxation rates $R_{2A}$ and $1/T_{2B}$, the offset frequency $\Delta$ (in Hz), exchange rate $k$, and the intensity of the applied RF field ($\omega_1$). The saturation rates can be determined from:

$$R_{rfB} = \pi\omega_1^2 g_B(\Delta) \tag{9}$$

$$\Omega = \frac{2\pi\Delta}{\omega_{loc}} \tag{10}$$

where $g_B(\Delta)$ is the normalized spectral lineshape of the semi-solid pool. $\omega_{loc}$ is a tissue-specific parameter which is dependent on the macromolecular lineshape, $T_{2B}$ and the tissue orientation, where the value for a super-Lorentzian lineshape averaged over all angles is $\sqrt{1/(15T_{2B}^2)}$ (19).

Saturation pulse shapes and the super-Lorentzian lineshape values were obtained from previously published code (21,22). Equations 4-8 can be combined into matrix form to produce an evolution matrix:

$$E = \begin{bmatrix} -R_{2A} & -2\pi\Delta & 0 & 0 & 0 \\ 2\pi\Delta & -R_{2A} & -\omega_1 & 0 & 0 \\ 0 & \omega_1 & -kM_0^B - R_{1A} & kM_0^A & 0 \\ 0 & 0 & kM_0^B & -R_{rfB} - kM_0^A - R_{1B} & \Omega R_{rfB} \\ 0 & 0 & 0 & \Omega R_{rfB} & -\Omega^2 R_{rfB} - \frac{1}{T_{1D}} \end{bmatrix} \tag{11}$$

with a time-dependent magnetization vector $M_t$ = [ $M_X^A$; $M_Y^A$; $M_Z^A$; $M_Z^B$; $\beta'$]. The magnetization at the next time step can be calculated as:

$$M_{t+\Delta t} = \exp(E\Delta t)\, M_t + E^{-1}(\exp(E\Delta t) - \mathrm{I}) \cdot B \tag{12}$$

where $\Delta t$ is the timestep, I is the identity matrix, and $B$ = [ 0; 0; $R_{1A}M_{Z(t=0)}^A$; $R_{1B}M_{Z(t=0)}^B$; 0].

Water Excitation

To accelerate the sequence for high resolution cortical imaging, it is beneficial to acquire multiple k-space lines per TR. Excitation pulses with flip angle α were simulated using a rotational matrix for the water pool, that was expanded to include a proportional saturation value for the bound pool:

$$R(\alpha) = \begin{bmatrix} 1 & 0 & 0 & 0 & 0 \\ 0 & cos(\alpha) & sin(\alpha) & 0 & 0 \\ 0 & -sin(\alpha) & cos(\alpha) & 0 & 0 \\ 0 & 0 & 0 & E_{rfB} & 0 \\ 0 & 0 & 0 & 0 & E_{rfD} \end{bmatrix} \tag{13}$$

where:

$$E_{rfB} = \exp\left(-R_{rfB_{exc}}\, t_{exc}\right) \tag{14}$$

$$E_{rfD} = \exp\left(-R_{rfB_{exc}}\, \Omega^2\, t_{exc}\right) = 0;\ (\Delta = 0) \tag{15}$$

The saturation parameters were calculated using the time of a square excitation pulse ($t_{exc}$).

The magnetization value following an excitation pulse is:

$$M_t = R(\alpha) M_{t-1} \tag{16}$$

Relaxation during the echo spacing was calculated using equations (11) and (12), using a single time step with $\Delta = 0$ and $\omega_1 = 0$.

**Methods:**

Imaging:

This study was approved by the Research Ethics Board of the Montreal Neurological Institute. Images were acquired on a 3 Tesla Siemens PrismaFit scanner using a 32-channel receive coil. Anatomical images were acquired using a custom MT-weighted RAGE sequence at 1 mm isotropic resolution, bandwidth = 250 Hz/px, sagittal acquisition, and a GRAPPA acceleration factor of 2 with 32 reference lines. The custom sequence contained several tuneable parameters relevant to ihMT contrast generation (**Figure 1**), including: saturation pulse length ($t_{pulse}$), saturation pulse gap ($t_{gap}$), number of saturation pulses ($N_{sat}$), number of saturation pulse trains per TR ($N_{burst}$), repetition times of saturation pulse bursts ($TR_{MT}$), and the turbo-factor. The sequence ran dummy scans for the first 5 seconds to drive the system into a pseudo-steady state. To improve SNR, a cartesian center-out radial fan beam k-space encoding was employed, as previously suggested (23,24), with two dummy echoes acquired at the beginning of each excitation train for turbo-factors>3. The matrix size of all MT-weighted volumes was 216x192x176 and the echo time (TE) was 2.65 ms. $B_1^+$ maps were acquired using a slice-selective preconditioning pulse with a turbo flash readout (25) with the following parameters: TR = 20 s, TE = 2.22 ms, 45 slices, FOV = 240x240 $mm^2$, matrix size = 96, distance factor = 20%, and 3 mm isotropic voxels.

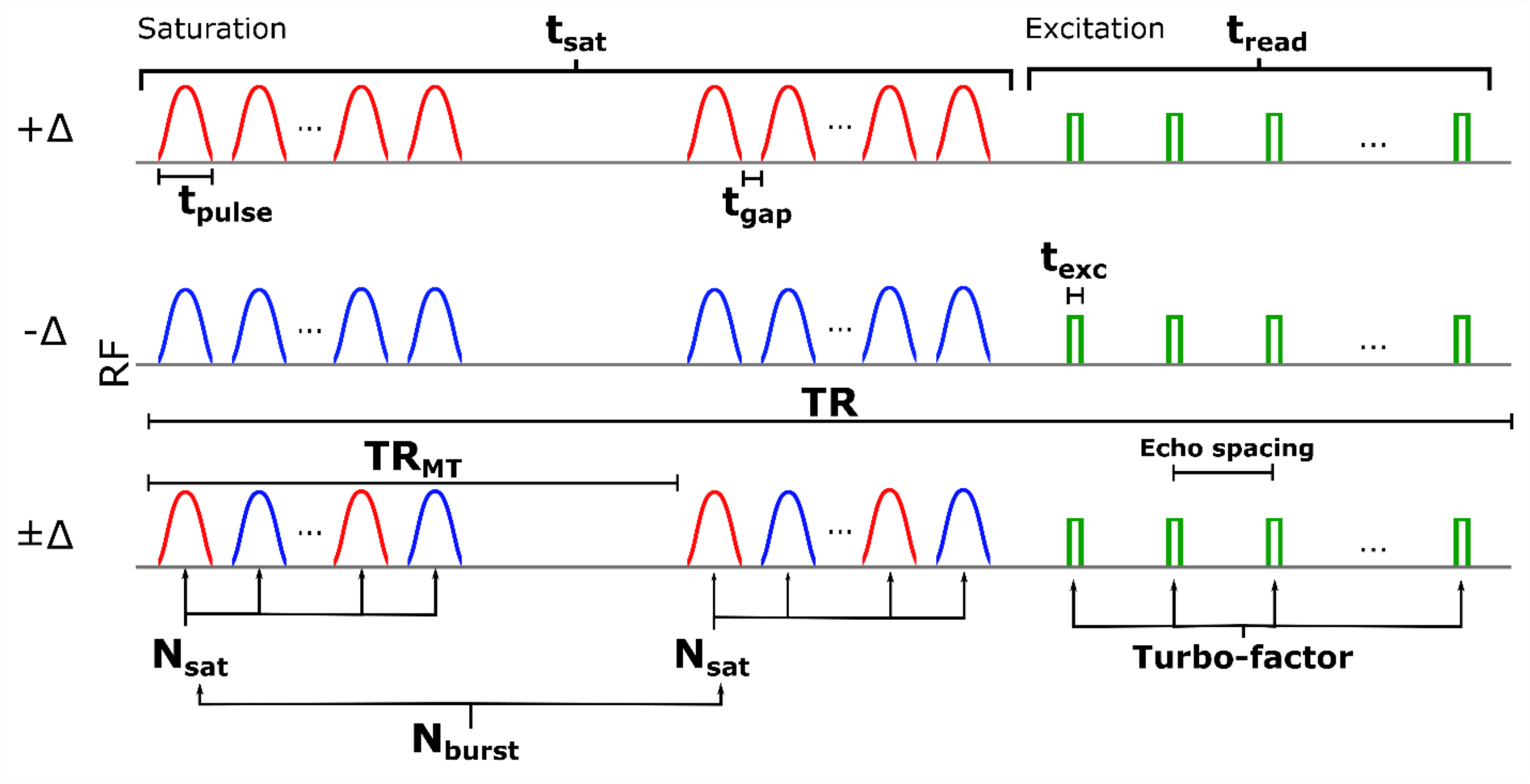

**Figure 1:** Representation of the MT-weighted RAGE sequence parameters that were modified in the cortical ihMT simulations. The saturation block with duration $t_{sat}$, and used pulse length ($t_{pulse}$), pulse gap ($t_{gap}$), number of pulses ($N_{sat}$), number of saturation pulse trains/TR ($N_{burst}$), and repetition time of pulse trains ($TR_{MT}$) as adjustable parameters. By setting $N_{burst} > 1$, it was possible to explore additional low duty-cycle saturation options. The turbo-factor was adjusted in the excitation block of duration $t_{read}$ and was combined with a cartesian center-out radial fan beam k-space ordering. The echo spacing was kept constant in the readout.

## *In Vivo* Imaging to Estimate Tissue Parameters:

Two healthy subjects (1 female, 29-31 years old) were imaged with three ihMT protocols to estimate tissue parameters for subsequent simulations. The protocols are presented in **Table 1** and were chosen to cover a large range of sequence parameters. All saturation pulses were applied using Hanning-shaped pulses with $t_{pulse}$ = 0.768 ms and $t_{gap}$ = 0.3 ms. Three MT-weighted images were acquired for each parameter set, one at each of the follow saturating frequencies: ±8 kHz (dual alternating), +8 kHz, and -8 kHz. A -100 Hz frequency shift was applied to all saturation pulses to minimize MT asymmetry (26). Echo spacing was 7.66 ms. MP2RAGE with compressed-sensing (27) was used to measure $T_{1,obs}$ and $M_0$ maps for $MT_{sat}$ calculation, where $T_{1,obs}$ is the estimation of $T_1$ using a single liquid pool. MP2RAGE parameters were: TR = 5 sec, flip angles = 4/5 degrees, inversion times = 940/2430 ms, turbo-factor = 175, 4.6x upsampling, jitter radius = 1.2, sampling density = 0.5, acquisition time of 4 mins 18 sec.

Equations for the calculation of $MT_{sat}$ with an arbitrary turbo-factor have been derived for use with dummy echoes (**Supplementary Material S.2**). $T_{1,obs}$, $M_0$ and $MT_{sat}$ maps were calculated with flip angles multiplied by the relative $B_1^+$ field to correct for $B_1^+$ inhomogeneity present in the excitation pulses. By not correcting for $B_1^+$ inhomogeneity in the saturation pulses, a range of cortical $MT_{sat}$ values was obtained with a spread of effective saturation pulse flip angles.

The background-denoised $T_1$-weighted UNI image from the MP2RAGE sequence was processed in FreeSurfer (28) (Version 7.2) with manual edits to correct the cortical segmentations by adding control points to include missing white matter and to remove dura mater from segmentations. Values from the calculated maps were sampled onto the mid-depth cortical surface. $MT_{sat}$ values that were corrected for excitation pulse $B_1^+$ inhomogeneity were extracted where $T_{1,obs}$ was between 1350-1450 ms, corresponding to the 1400 ms value used in the subsequent simulations. This corresponds to the $T_{1,obs}$ of cortical GM when measured with MP2RAGE (29).

To extract appropriate tissue parameters, all three protocols in **Table 1** were simulated with a range of parameters. The tissue parameters that presented the lowest mean squared difference between the simulation data and the pooled subject data were selected as the best fit. Since the lineshape is simulated as being symmetric ($g_B(+\Delta) = g_B(-\Delta)$), the data was pooled for fitting the $+\Delta$ and $-\Delta$ $MT_{sat}$. The simulated parameter range included: k from 15-50 $s^{-1}$, $T_{2A}$ from 20-90 ms, $T_{1D}$ from 0.5 to 6 ms, $T_{2B}$ from 8-12 μs, $M_0^B$ from 0.0475-0.07, $R_{1B}$ from 0.25 to 1 $s^{-1}$, and $B_{1rms}$ from 0-18 μT (ranges obtained to cover previous works (18,30,31)).

| Parameters | Protocol 1 | Protocol 2 | Protocol 3 |
| --- | --- | --- | --- |
| TR (ms) | 120 | 1140 | 3000 |
| Flip Angle (deg) | 5 | 7 | 11 |
| Turbo-factor | 8 | 80 | 200 |
| $N_{sat}$ | 6 | 6 | 10 |
| $N_{burst}$ | 1 | 9 | 10 |
| $TR_{MT}$ (ms) | NA | 60 | 90 |
| Sat Pulse Flip Angle (deg) | 134 | 136 | 157 |
| $B_{1rms,pulse}$ (μT) | 13.94 | 14.15 | 16.33 |
| Acquisition Time | 4:32 (x3) | 4:20 (x3) | 4:35 (x3) |

**Table 1**: Acquisition parameters for estimation of the cortical tissue parameters in the two-pool tissue model with dipolar order.

Simulating Protocols for Optimal ihMT SNR:

MRI is constrained by a maximum amount of energy per unit time, which is proportional to the square of the applied RF field over a given time ($B_{1rms}^2$). To remain within these limitations, the peak $B_1^+$ was limited to 28 μT to respect hardware constraints, with the long-term SAR limit set to 3 W/kg. The deposited power can be calculated using:

$$P = c\omega_0^2 {B_1}^2 \tag{17}$$

where $\omega_0$ is the frequency ($\gamma B_0$), and c is an empirical factor derived to match SAR values for a subject weighing 60 kg, equalling $1.44 \cdot 10^{-3}$. For simulations, the peak $B_1^+$ of the saturation pulses was scaled to the maximum allowable for each set of sequence parameters. This led to some protocols having a lower $B_{1rms,TR}$ in cases where the peak $B_1^+$ limit would have been exceeded.

The saturation pulses were selected to be far off-resonance (8 kHz), with minimal direct-saturation expected, even when accounting for the increased bandwidth of short pulses ($t \geq 0.768$ ms). This permitted the removal of spoiling between saturation pulses in favour of a faster switch time in the dual saturation sequence to increase ihMT SNR (32), however this may not work with shorter pulses, or with other pulse shapes with increased bandwidths (10,15). A 1.4 ms spoiler gradient was used at the end of the MT-prep block to dephase any transverse magnetization prior to the RAGE readout. In the simulations, the transverse magnetization was completely spoiled ($M_x = M_y = 0$) following the MT-prep block and before each excitation pulse. The custom MRI sequence utilized RF and gradient spoiling during the excitation block.

A decrease in magnetization over the excitation train can lead to an artificial increase in calculated MT metrics. To extract an equivalent image intensity value from a single simulated voxel, a sampling table was generated for each turbo-factor simulated. Once a pseudo-steady state was reached, the magnitude of the transverse magnetization following each excitation pulse was used to fill the simulated k-space. Each k-space point has a differential weighting towards a voxel's intensity value in a brain image. To derive a k-space weighting, the MNI-152 atlas (33) was chosen as a template brain segmentation (brain = 1, else = 0). The Fourier transform of the atlas brain mask was used as a multiplicative weighting factor that was applied to the simulated k-space data. The resulting weighted k-space values were inverse Fourier transformed, and the mean intensity within the brain mask was extracted as the resulting value for the simulated parameter set. A flow chart documenting this process for two turbo-factors can be found in **Supplementary Materials Figure S2**.

Simulations were run for single and dual off-resonance MT-weighting, as well as no MT-weighting for MTR (MT ratio) calculation. To view the impact of the different low duty-cycle MT approaches, simulations were separated into two groups: $N_{burst}$=1 and $N_{burst}$>1. Only parameter combinations that permitted a volume to be acquired within 4.5 minutes were simulated.

$MT_{sat}$ was calculated for each MT-weighting, and $ihMT_{sat}$ values were calculated as $MT_{sat,dual} – (MT_{sat,pos} + MT_{sat,neg})/2$. While we assumed the lineshape to be equivalent, and thus simulated $MT_{sat,pos} = MT_{sat,neg}$, this equation was used to be consistent with the *in vivo* SNR expectations.

Since ihMT varies linearly with myelin content, this study used the resulting SNR to compare the efficacy of the parameter sets. SNR was calculated for each simulated protocol, assuming additive white gaussian noise with a constant standard deviation across images of 0.0005 (~SNR of 70). Details on SNR calculation are presented in **Supplementary Materials S.4**.

Simulations – ihMT PSF Comparison:

| **Parameters** | **$N_{burst}$ = 1** | **$N_{burst}$ > 1** |
|---|---|---|
| Turbo-factor | 1 - 90 | 16 - 200 |
| TR (ms) | 15 - 1140 | 200 - 4000 |
| Flip Angle (deg) | 4 - 13 | 5 - 11 |
| $N_{sat}$ | 2 - 20 | 4 - 20 |
| $t_{pulse}$ (ms) | 0.768, 1.024, 1.28, 2.048 | 0.768, 1.024 |
| $N_{burst}$ | 1 | 2 - 14 |
| $TR_{MT}$ (ms) | NA | 40 - 150 |
| Combinations within 4.5 min scan time | 7947 | 207366 |

**Table 2:** The range of sequence parameters that were simulated for their impact on ihMT SNR. Parameters that were fixed included $t_{gap}$ = 0.3 ms, and Δ = 8 kHz.

The cartesian center-out radial fan beam RAGE readout is used to increase SNR, but this can have a large impact on the effective resolution. As the readout parameters were modified to accommodate different saturation schemes within a fixed scan time, the impact of increasing the turbo-factor on the effective resolution was investigated for the best SNR protocols over a range of turbo-factors. The investigated sequence parameters are listed in **Table 2**. The sampling table was generated for each turbo-factor protocol and filled with the transverse magnetization values from the numerical simulations. The inverse 2D Fourier transform was taken of the simulated acquisition matrix, and the result was upsampled by a factor of 100. A 1-D line was extracted from the center of the PSF to view the in-plane PSF, i.e., perpendicular to the readout direction. Three protocols providing the best ihMT SNR with turbo-factors of 10, 80 and 200 were collected in one healthy adult (female, 32 years old), with the acquisition parameters listed in **Table 3**, however the 200 turbo-factor protocol was acquired with $N_{burst}$ = 11 and $TR_{MT}$ = 60 ms.

| **Turbo-factor** | **10** | **48** | **80** | **120** | **160** | **200** |
|---|---|---|---|---|---|---|
| TR (ms) | 140 | 750 | 1250 | 1750 | 2300 | 2900 |
| Flip Angle (deg) | 5 | 7 | 9 | 11 | 11 | 11 |
| $N_{sat}$ | 6 | 4 | 6 | 6 | 6 | 8 |
| $N_{burst}$ | 1 | 6 | 7 | 10 | 13 | 12 |
| $TR_{MT}$ (ms) | NA | 40 | 40 | 40 | 40 | 40 |
| Sat. Pulse Flip Angle (deg) | 145 | 163 | 162 | 159 | 160 | 162 |
| $B_{1rms,pulse}$ (μT) | 15.1 | 16.97 | 16.91 | 16.58 | 16.67 | 16.88 |
| $B_{1rms,sat}$ (μT) | 12.52 | 4.70 | 5.74 | 5.63 | 5.65 | 6.62 |
| $B_{1rms,TR}$ (μT) | 2.75 | 2.69 | 2.76 | 2.76 | 2.76 | 2.76 |
| Duty Cycle (sat period) | 0.7191 | 0.0902 | 0.1309 | 0.1258 | 0.1232 | 0.1644 |
| Duty Cycle (TR) | 0.0401 | 0.0310 | 0.0322 | 0.0332 | 0.0330 | 0.0323 |

**Table 3:** Sequence parameters as derived from simulations to cover a range of turbo-factors that provide optimal ihMT SNR.

*In Vivo* Experiment – Cortical Maps:

A single $ihMT_{sat}$ protocol was collected in a group of 25 subjects aged 19-40 years (mean: 29 ± 7 years, 12 female) to demonstrate the potential of 1 mm isotropic $ihMT_{sat}$ to study cortical microstructure. A protocol was selected with a higher effective resolution at the cost of SNR, to examine if sufficient SNR can be attained at 1 mm nominal resolution to assess spatial patterns in cortical myelination. The protocol values were: TR = 100 ms, TE = 2.8 ms, flip angle = 6 deg, FOV = 256x256x176, turbo-factor = 8. The saturation parameters included $t_{pulse}$ = 0.768 ms, $t_{gap}$ = 0.3 ms, Δ = 8 kHz, $B_{1rms,pulse}$ = 15.0 μT, and $N_{sat}$ = 4 and $N_{burst}$ =1. Surfaces were generated as in the tissue parameter estimation section, using FreeSurfer (28) with manual edits. Values sampled onto the mid-depth cortical surface were smoothed and displayed in MATLAB (version R2021b) using Surfstat (http://www.math.mcgill.ca/keith/surfstat/).

The MATLAB simulation code and the sample data that were used for fitting tissue parameters are available at: https://github.com/TardifLab/OptimizeIHMTimaging.

**Results:**

*In Vivo* Imaging to Fit Tissue Parameters:

The acquired $MT_{sat}$ values from the cortex of two subjects are plotted in **Figure 2**. The tissue parameters that provided the lowest mean squared error to the extracted cortical values are: $M_0^B/M_0^A$ = 0.071; $R_{1obs}$ = 0.714 $s^{-1}$; k = 50 $s^{-1}$; $T_{2A}$ = 50 ms; $T_{1D}$ = 0.75 ms; $R_{1B}$ = 0.25 $s^{-1}$; $T_{2B}$ = 11.5 μs. Quantitative MT fitting is challenging due to its high dimensionality; we therefore acknowledge that this may have fit a local minimum. As such, these are not intended to accurately reflect the underlying biophysical model, but rather serve as a guide for simulations within this range of sequence parameters.

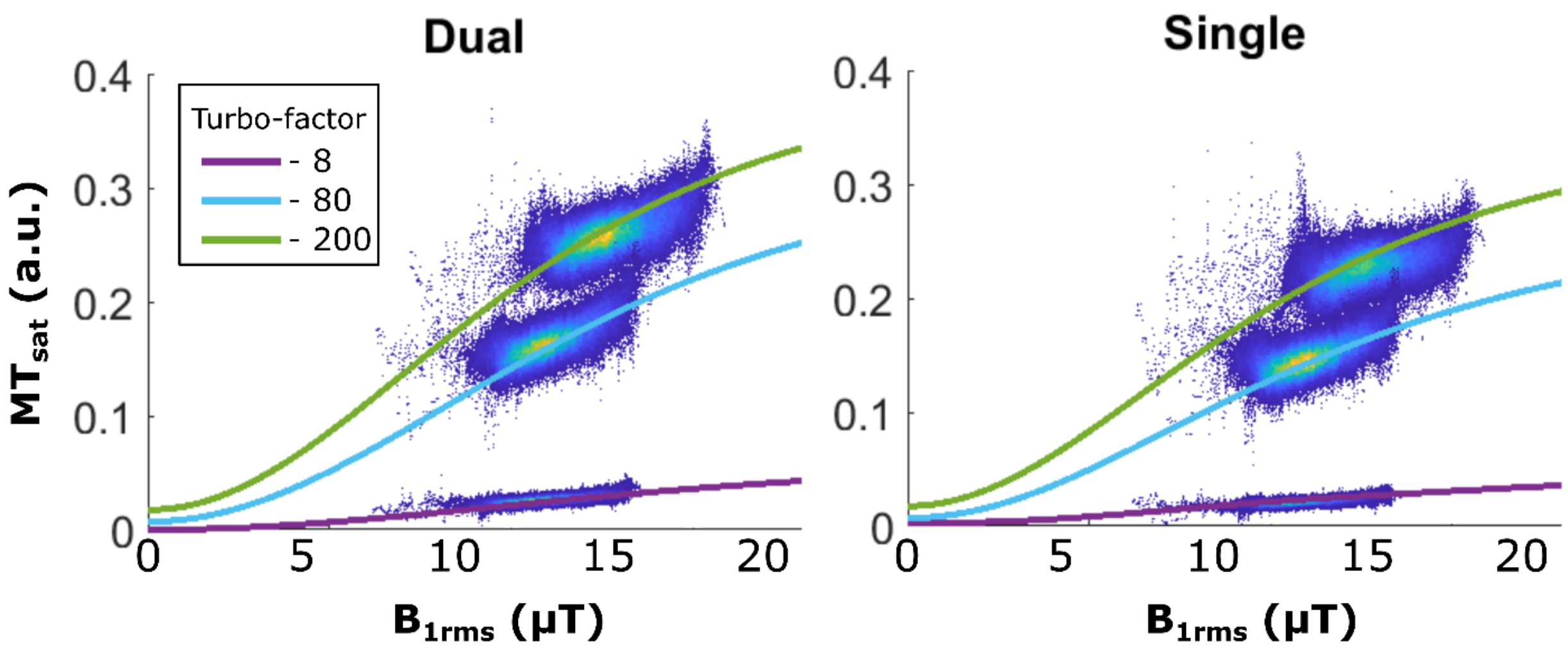

**Figure 2:** A heat scatter plot shows the pooled $MT_{sat}$ data from two subjects from cortical regions where $T_{1,obs}$ was between 1350 and 1450 ms, against the $B_{1rms}$ of the saturation pulses. $MT_{sat}$ values were generated from three separate protocols, each with ±8 kHz (dual alternating), and with two single sided saturation schemes (+8 kHz, and -8 kHz). Plotted lines present the fit result from the tissue parameters that provided the lowest mean squared error. The simulations provide good fit to both the dual and single saturation data across all three protocols.

Simulating Protocols for Optimal ihMT SNR:

The resulting ihMTR SNR was also calculated (not shown) and was found to follow the same SNR trends as $ihMT_{sat}$. Peak ihMTR SNR was 2.373 and 7.824 for $N_{burst}$=1 and $N_{burst}$>1 respectively. These values were 2.354 and 7.822 for $ihMT_{sat}$, representing a 0.8% and 0.03% difference between the approaches respectively.

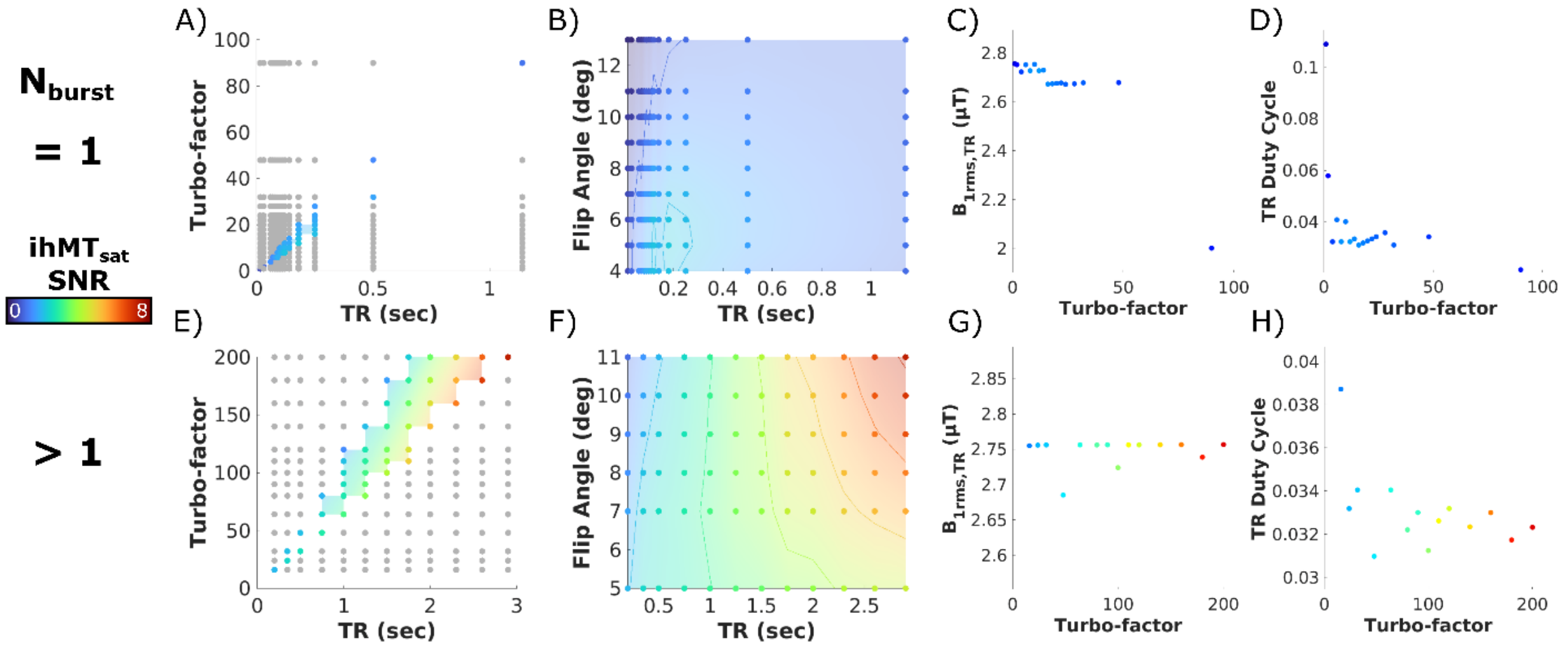

**Figure 3:** Contour plots derived from simulations present how excitation parameters impact $ihMT_{sat}$ SNR. Each point represents the greatest simulated SNR that was possible with only the axis variables fixed, and the remainder free. Grey points represent a parameter combination that did not fit within the 4.5 min/volume constraint. Colour values are interpolated between simulated parameter combinations to aid in visualizing trends. Plots A and E highlight the trade-off between TR and turbo-factor needed for fast, high-resolution imaging. Plots B and F shows the ideal excitation flip angle to use. Plots C and G present the $B_{1rms,TR}$ for all RF pulses that provide the best $ihMT_{sat}$ SNR. Plots D and H highlight the increased SNR can be achieved with lower RF duty cycles when calculated over the whole TR.

---

The impact of the excitation block on $ihMT_{sat}$ SNR are displayed in **Figure 3**. Plots A and E demonstrate combinations of TR and turbo-factor that are necessary for high resolution imaging with the 4.5 min/volume constraint. In plot B, $ihMT_{sat}$ SNR is optimized for flip angles of 4-6 degrees and a TR between 80-200 ms with $N_{burst}$=1. SNR increases with increasing TR and flip angles with $N_{burst}$>1 as displayed in plot F. Plots C and G present the $B_{1rms,TR}$ providing the greatest $ihMT_{sat}$ SNR, which was close to 2.75 μT and decreased slightly with increasing turbo-factor with $N_{burst}$=1. The consistency in the plotted $B_{1rms,TR}$ for $N_{burst}$>1 suggests that protocols that produces the greatest SNR were close to the power constraints. Plots D and H shows the corresponding RF duty cycles over the TR, with optimal SNR being attained with duty-cycles close to 3% with the study constraints. With $N_{burst}$>1, large increases in $ihMT_{sat}$ SNR are observed, without large differences in duty cycle or $B_{1rms,TR}$.

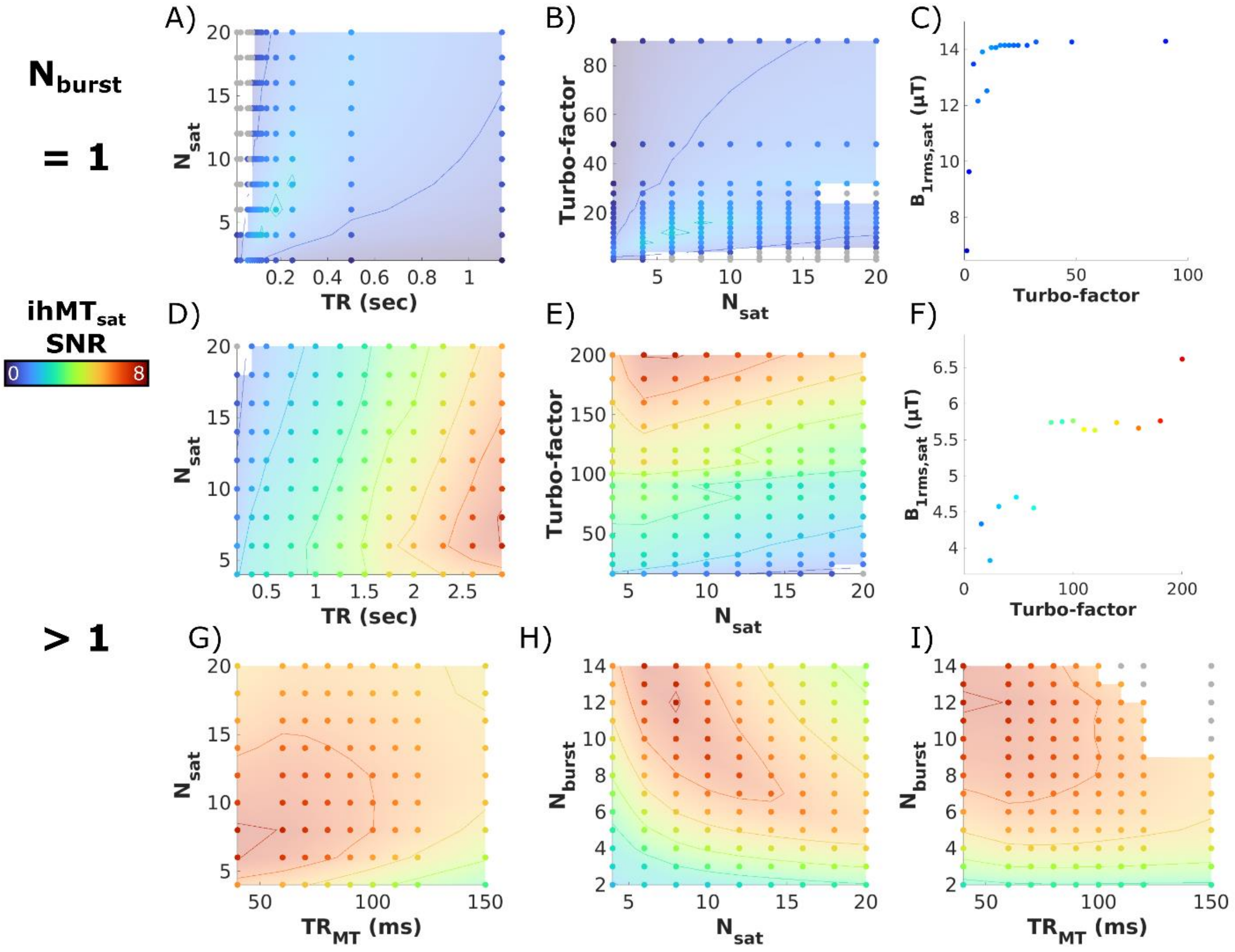

**Figure 4:** Simulations results are plotted to present the impact on saturation parameters on $ihMT_{sat}$ SNR. Each point represents the best simulated SNR that was possible with only the axis variables fixed, and the remainder free. Grey points represent a parameter combination that did not fit within the 4.5 min/volume constraint. Plots A, B, D and E highlight the differential optimization that is used by increasing $N_{burst}$. Plots C and F present the impact of turbo-factor on the $B_{1rms,sat}$ that provides the highest $ihMT_{sat}$ SNR. A decrease in $B_{1rms,sat}$ is observed with $N_{burst}$ > 1, with higher values producing greater $ihMT_{sat}$ SNR. Plots G, H and I present the impact of the additional parameters that are available with $N_{burst}$ > 1, for optimizing $ihMT_{sat}$ SNR.

The impact of the saturation parameters on $ihMT_{sat}$ SNR are displayed in **Figure 4**. Plots A, B, D, E, G and H suggest that the ideal number of saturation pulses is between 6-8 in each burst. A region of optimal parameters was identified for the protocols with $N_{burst}$=1, with turbo-factors between 8 and 16, and TR between 80 and 200 ms. With $N_{burst}$>1, $ihMT_{sat}$ SNR improved with increasing TR and turbofactor. Plots C and F highlight the increase in SNR but a decrease in $B_{1rms,sat}$, when $N_{burst}$>1. Increases in $B_{1rms,sat}$ in the $N_{burst}$>1 protocols produced greater ihMT SNR. For the protocols with $N_{burst}$>1, an optimal saturation procedure consists of a $TR_{MT}$ between 40-60 ms, and a maximized $N_{burst}$ for the TR.

<u>Simulations – ihMT Point Spread Function Comparison</u>:

The PSFs from the parameter sets that provided the highest ihMT SNR for a range of turbo-factors are presented in **Figure 5**, with parameters outlined in **Table 3**. All PSFs were normalized to the largest value of all PSFs for comparison. The protocol with $N_{burst}$=1 and a turbo-factor of 10 presented the narrowest in-plane point-spread function. The approaches with $N_{burst}$>1 and turbo-factors of 48 and 80 had a minor increase in the full width at half maximum (FWHM), and decrease in peak height. Protocols with turbo-factors of 120, 160 and 200 had broad PSFs, with FWHMs exceeding 2 mm. The ihMTR SNR was plotted as a function of the PSF FWHM in **Figure 5B**. When compared to the approach with a low turbo-factor and $N_{burst}$=1, the other approaches provided improved simulated ihMTR SNR than what would be expected by a change in voxel size alone (solid blue line). The increase in blurring at high turbo-factors can be observed in **Figure 5C**, with minimal change in blurring between turbo-factors of 10 and 80. While the acquired data for turbo-factor = 200 uses different saturation parameters from those presented in Table 3, the readout parameters that largely drive the blurring effect with k-space ordering are the same. **Figure 5D** illustrates the dual $MT_{sat}$ maps that also present increased blurring when acquired with larger turbo-factors.

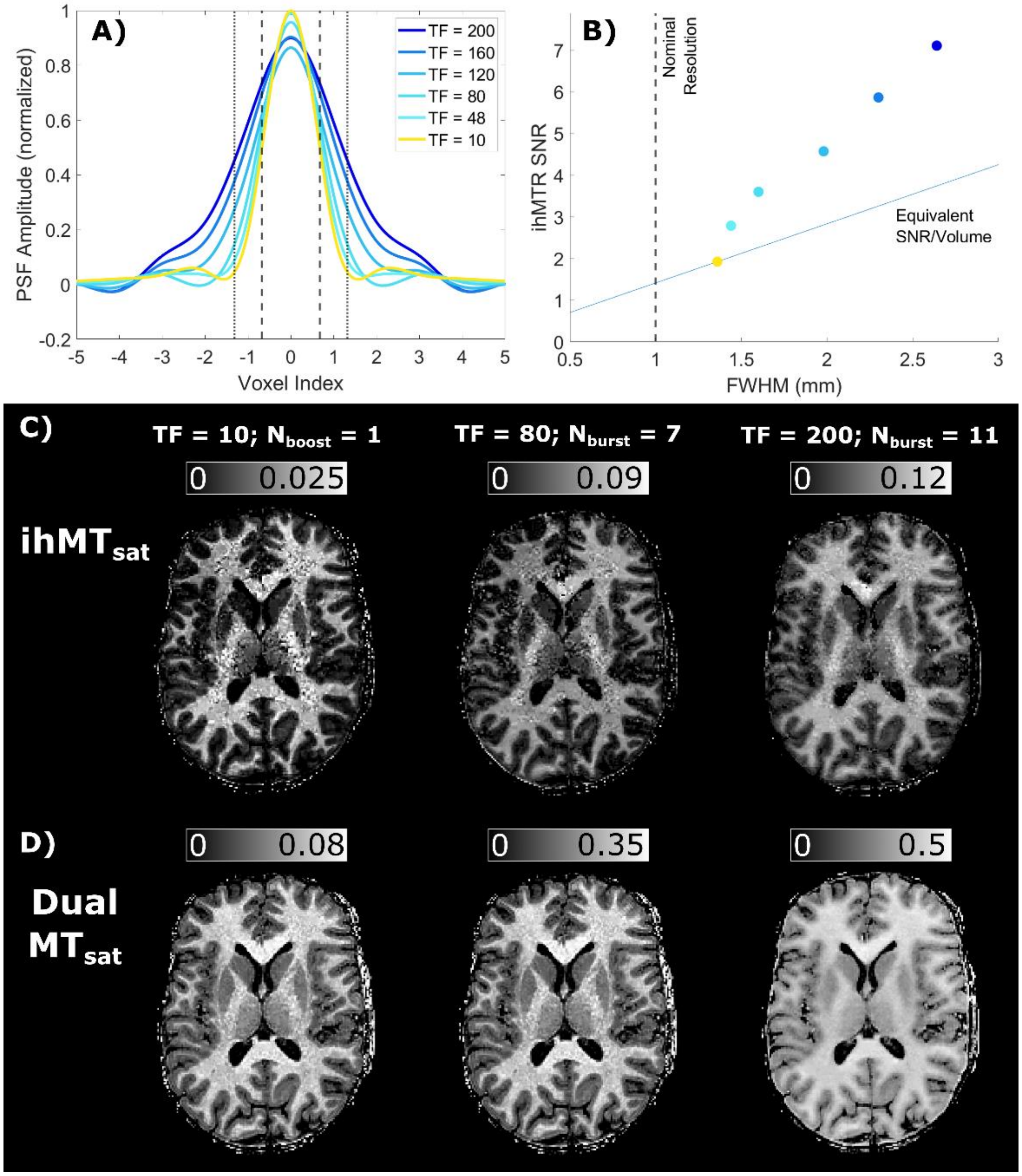

**Figure 5:** A point spread function analysis was used to visualize the impact of turbo-factor (TF) and $N_{burst}$ on effective resolution. A) The narrowest FWHM of the plotted PSFs is obtained with TF = 10 and $N_{burst}$ = 1, with an increase in FWHM at much larger TF. The dashed and dotted lines mark the smallest and largest FWHM plotted, respectively. B) A comparison of the effective resolution (FWHM) and the ihMTR SNR. The dashed line marks the nominal resolution. The blue line uses the $N_{burst}$ = 1 point as the reference and tracks how SNR would change by modifying the nominal resolution. By increasing $N_{burst}$, an SNR increase is observed that is larger than what would be achieved by increasing voxel size, as all points are above the blue line. C) $ihMT_{sat}$ axial slices corroborates the simulation results, as there is a minimal increase in blurring when increasing TF from 10 to 80, with more pronounced blurring at TF = 200. D) Corresponding dual $MT_{sat}$ maps also demonstrate increased blurring with larger turbo-factor. The protocols details are in Table 3, but the TF = 200 uses $N_{burst}$ = 11, and $TR_{MT}$ = 60 ms.

*In Vivo* Experiment – Cortical Maps:

A 25-subject group average, as well as two subjects' $ihMT_{sat}$ values are presented on an inflated FreeSurfer average mid-depth surface in **Figure 6**. Spatial patterns that match the myeloarchitecture of the cortex (34) can be observed on the average surface, with primary motor and somatosensory cortices appearing brighter than neighbouring association cortices. Simulations suggest that an improvement in SNR of 12% can be achieved by using the optimized parameters in **Table 3** for the turbo-factor = 10 protocol.

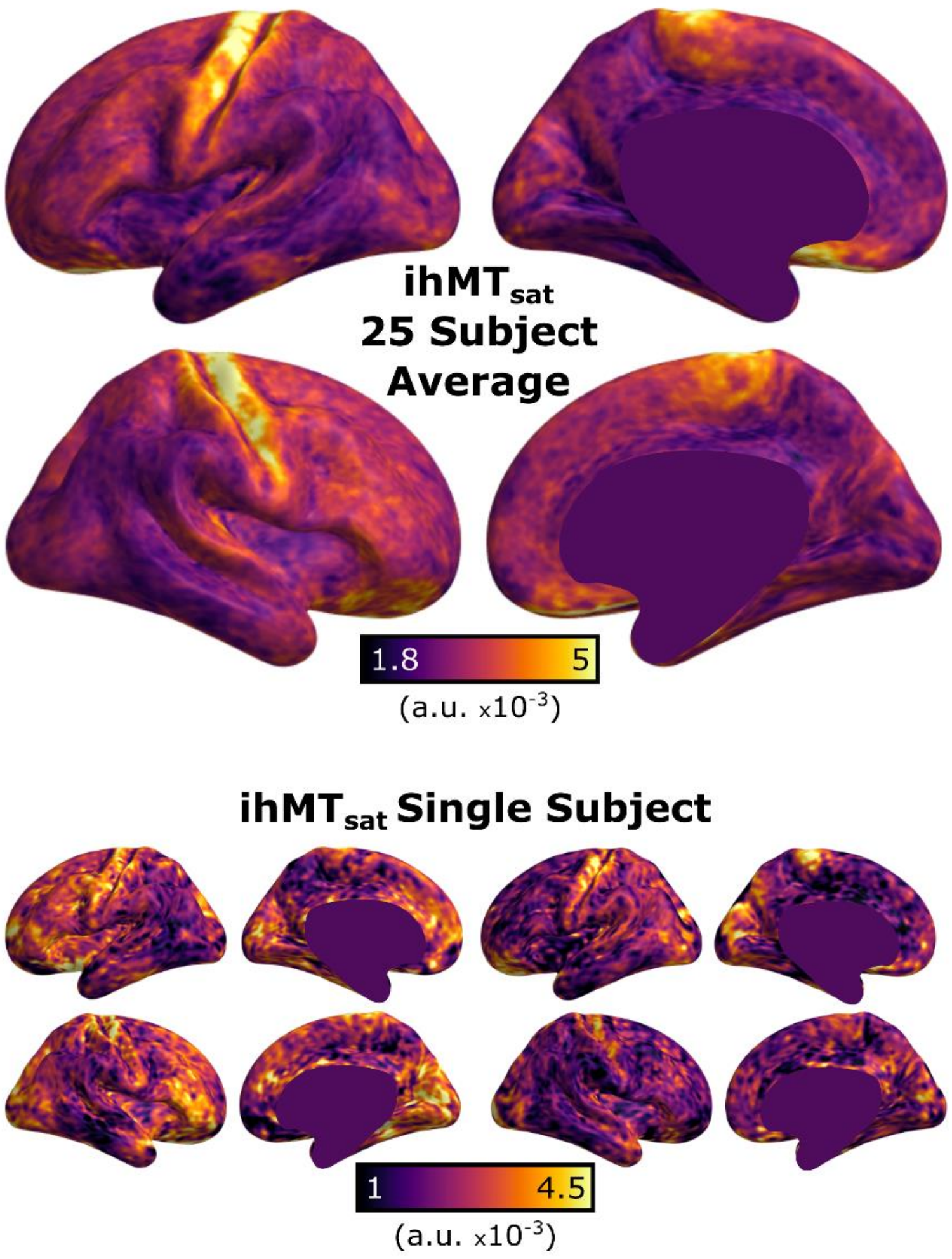

**Figure 6:** Top) Group average cortical $ihMT_{sat}$ map acquired at 1 mm nominal resolution in 25 subjects. Maps were smoothed with a 3 mm Gaussian kernel before averaging and displayed on an inflated surface. Despite the lower predicted SNR with $N_{burst}$ = 1, well known spatial patterns in cortical myelination can be observed. Primary cortical areas display prominently from the surrounding association regions that contain lower myelin content. Bottom) Representative cortical $ihMT_{sat}$ map from two subjects that were smoothed with a 5 mm gaussian kernel.

**Discussion:**

In this study, saturation and readout parameters were examined for their impact on ihMT SNR and effective resolution in the pursuit of generating an optimized protocol for high-resolution cortical imaging in a feasible scan time. It was found that higher SNR values are achieved by increasing $N_{burst}$ while maintaining TR duty cycles around 3%, saturation duty cycles >10% and $B_{1rms,TR}$ values close to 2.75 μT. An optimal parameter region was found for $N_{burst}$=1. With $N_{burst}$>1, there was a clear trend of increasing ihMT SNR with increasing turbo-factor, TR and $N_{burst}$. For the time constraint of 4.5 mins per acquired volume, we found an ideal combination of saturation parameters that uses a $TR_{MT}$ of 40 ms and 6-8 saturation pulses. The PSF analysis demonstrated that the increased SNR with higher turbo-factors is associated with a lower effective resolution. The simulation results suggest that high SNR and resolution can be achieved with turbo-factors at or below 80 with $N_{burst}$ > 1. Finally, as a proof of concept, we presented a 1-mm isotropic cortical $ihMT_{sat}$ group average map consisting of 25 subjects.

The partial quantitative MT experiment fixed the values of $M_0^A$ and $R_{1obs}$ with the fitted values within the range of previously reported GM values (18,30,31). $T_{2A}$ variations (20-100 ms) had minimal impact on our fit, likely due to the large offset frequency of the saturation pulses limiting direct saturation. With only one offset frequency used, there was also limited sensitivity to $T_{2B}$. While $R_{1B}$ is typically fixed to 1 $s^{-1}$, we observed a small improvement in fit with $R_{1B}$ = 0.25 $s^{-1}$. The fitted $T_{1D}$ value (0.75 ms) was lower than a previous study that reported $T_{1D}$ in the GM of 5.9 ms (35). This value was consistent with previous fitting approaches using conventional MT data and a single bound pool model (6,19). A fitted sub-millisecond $T_{1D}$ appears to result from fitting conventional MT data, with the decrease possibly arising from the high correlation with $M_0^B$. When this variable is removed from the fitting process by acquiring images specifically sensitive to $T_{1D}$, more accurate decay values can be obtained (35).

$T_{1D}$ drives the contrast observed in ihMT maps, with myelin lipids possessing a significantly longer $T_{1D}$ than many cellular proteins (9,36). This is presumably due to their structural arrangement that limits mobility and increases interaction times for dipolar order (10). Shorter switching times will provide higher SNR, but this comes at the cost of specificity as there is less filtering towards the longer $T_{1D}$ components of myelin lipids. It has been shown that ihMT maps are less correlated with myelin when shorter switch times are used (8). Recently, MT models have expanded to include multiple $T_{1D}$ components to increase specificity of imaging towards longer $T_{1D}$ signal components (8,37,38). Other work has demonstrated that $T_{1D}$ is not a stable tissue parameter, as it changes with temperature (9,39), and with the molecular orientation to $B_0$ (40). While recent work has used $T_{1D}$-filtering to increase myelin specificity, a lower switch time as was used in this study, is likely necessary for high-resolution cortical imaging due to the low myelin content and thus low ihMT SNR.

The $B_{1rms,TR}$ for the simulations was calculated to be around 2.75 μT, which is slightly lower than the 3 μT reference used in previous studies (15,24). We aimed to underestimate the long-term $B_1^+$ limits by using a lower weight in the power calculation to derive protocols that should meet SAR demands in most subjects. Previously, an RF duty-cycle of 5% calculated over the saturation block was demonstrated to provide the greatest ihMTR using 5 ms pulses and cosine-modulated dual frequency pulses (14). This saturation strategy was subsequently used for 3D ihMT imaging (12,24). When using a centric-encoded RAGE readout, our simulations suggest that increased ihMT SNR is possible using saturation RF duty cycles over 10%, provided that the applied RF pulses are maintained at near maximal power given the hardware constraints. The duty cycle over the entire TR is ~3%, permitting the saturation parameters to fall within SAR constraints. Our results thus agree with previous work highlighting that increases in ihMT SNR is driven by high power saturation pulses, which is facilitated by low duty-cycles over the TR on clinical scanners.

The discrete sampling nature of the sequence parameters may have hidden more optimal parameter combinations from our search. Compared to previous studies, the current work included additional time constraints and studied the impact of increasing the turbo-factor on resulting ihMT values. Our simulations used short saturation pulses with a 0.3 ms gap that have previously been shown to produce a strong ihMT effect, and to provide increased specificity over continuous cosine-modulated pulses (4,6,32). Our results corroborate previous findings suggesting that shorter pulses with high peak $B_1^+$ and increased RF irradiation duration increase ihMTR (4,41). This study supports the notion that optimal ihMT SNR with weak $T_{1D}$ filtering is achieved through low duty-cycle approaches (14), which is enhanced by increasing $N_{burst}$ with roughly 6 pulses per burst combined with a $TR_{MT}$ of 40-100 ms (7,41).

However, we demonstrate that the highest SNR approach may be unsuitable for cortical imaging when combined with an increased turbo-factor due to the resulting lower effective resolution. The effective resolution decrease could be lessened by using a linear k-space encoding scheme. The effect of which will depend on the turbo-factor, but this will result in significantly lower SNR for the long turbo-factor protocols if constant flip angles are used in the excitation block. This investigation was restricted to RAGE readouts, but other readouts could be used. Recently, it was demonstrated that a fast spin echo (FSE) readout can provide a significant increase in ihMT SNR. However, an optimization of this protocol from 2.4x2.4x2.8 $mm^3$ to high-resolution imaging has not been done.

The simulation results suggest that increasing the number of saturation pulse trains in each TR is an effective way to increase ihMT SNR. This study limited sequence parameter options to acquire each volume in 4.5 mins or less. To achieve that, turbo-factors had to be increased to permit long saturation blocks with high $B_{1rms}$ saturation pulses. The change in the longitudinal magnetization over the long excitation trains, combined with the center-out radial fan beam encoding, lead to an increased FWHM of the PSF. This presents a trade-off between resolution and SNR. The axial slices in Figure 5 present increased noise towards the center of the brain, which is expected based on the sensitivity profiles of the receive coils (43). Thus, if a lower resolution is deemed sufficient to study the WM, then a higher turbo-factor would be appropriate to boost SNR. While optimizations were performed to maximize SNR, sequence parameters could also be selected to reduce sensitivity to $B_1^+$ inhomogeneity (44), which may be of particular interest at ultra-high fields.

Cortical ihMT imaging was previously demonstrated at 1.6 mm nominal resolution (12) using a protocol that was more aligned with the higher SNR approaches found here through simulations and demonstrated in a single subject (**Figure 5**). Comparing the single subject images from Munsch et al. with the images corresponding to a turbo-factor of 80 in Figure 5C and 5D, the lower SNR in the present study is apparent due to the 4-fold smaller voxel size (1.6 vs 1 $mm^3$ isotropic). For cortical mapping, the present study employed a lower SNR approach to obtain higher effective resolution to look at intracortical variation at the group level. If we compare the group average maps between the studies, the 1 $mm^3$ isotropic map demonstrates strong $ihMT_{sat}$ increases in primary motor and sensory cortices with lower values and minor fluctuations in other cortical areas. The 1.6 mm$^3$ maps exhibit larger variations in other association cortices, such as in supplemental and pre- motor areas. These differences could be due to differences in SNR but could also arise from partial volume effects that stem from the different resolutions utilized. As the cortex varies from 2-4 mm in thickness, a protocol with high effective resolution is necessary to rule out contributions from neighbouring white matter and cerebral spinal fluid. Future studies could increase SNR by using the optimized protocols for turbo-factors of 48 or 80, allowing visualization of more subtle cortical patterns with limited impact on effective resolution.

**Conclusions:**

This study builds on previous work highlighting the applicability of ihMT imaging for studying cortical myelination. We optimized sequence parameters for a given imaging time to maximize ihMT SNR, permitting the acquisition of 1 mm isotropic ihMT maps in less than 20 minutes. While other metrics have demonstrated good correlations with myelin content in healthy grey matter, ihMT imaging will be advantageous for studying myelin changes in neurological diseases where changes in myelin content could be confounded with inflammation and protein accumulation. For instance, high-resolution ihMT imaging would be ideal for assessing myelination patterns and lesion load in the cerebral cortex in multiple sclerosis.

**Data Availability Statement:**

The simulation code and the sample data that was used for fitting tissue parameters are available at: https://github.com/TardifLab/OptimizeIHMTimaging.

**Acknowledgements:**

This project has been made possible by the Brain Canada Foundation, through the Canada Brain Research Fund, with the financial support of Health Canada and the Natural Sciences and Engineering Research Council of Canada (C.R., J.C., G.B.P., and C.T.), the Fonds de recherche du Québec— Santé (C.T.), Healthy Brains for Healthy Lives (C.R., and C.T.), the Campus Alberta Innovates Program (G.B.P.).

**Supplementary Information:**

## S.1 Two-pool model with dipolar order

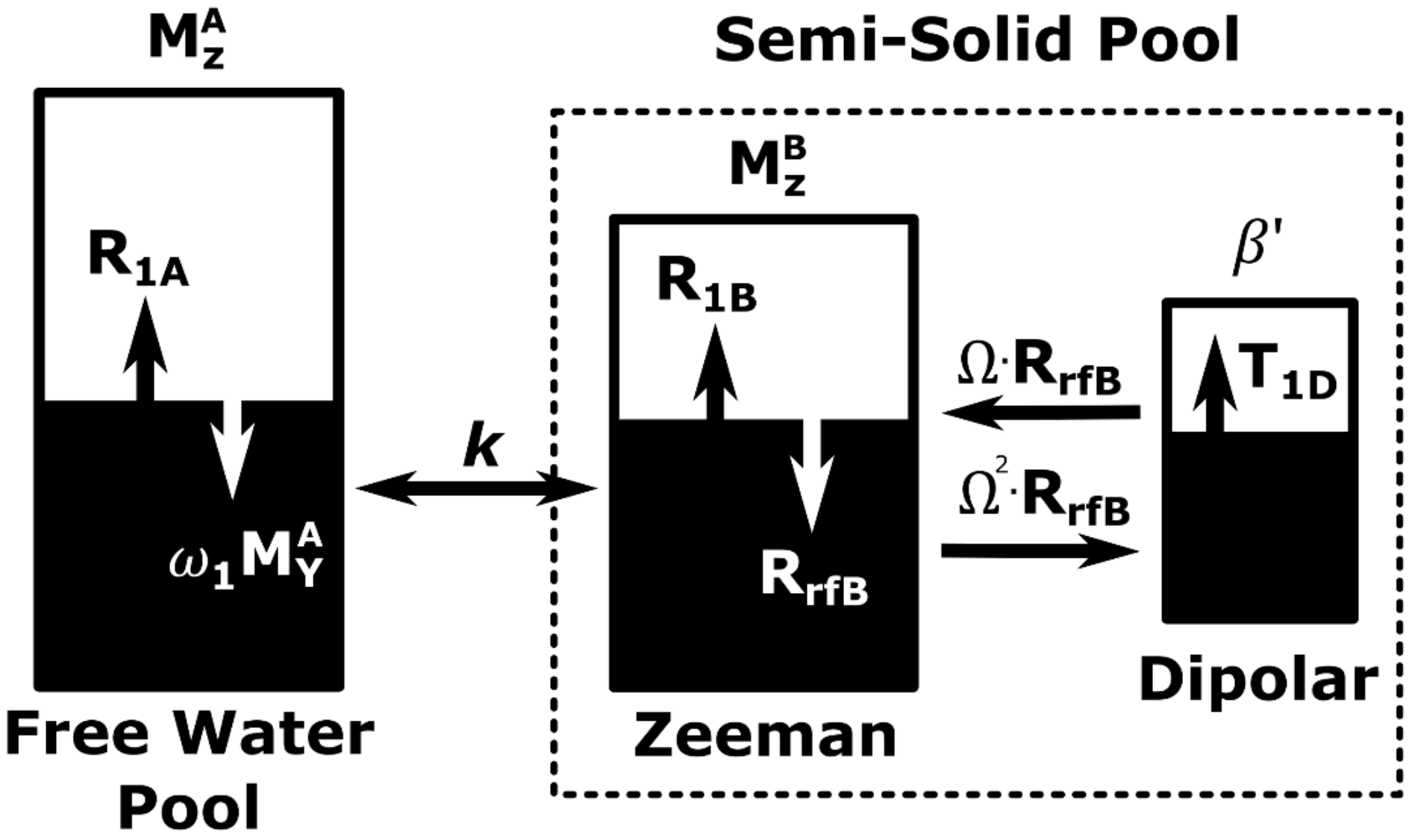

**Figure S1:** Illustration of the two-pool model with dipolar order. Image adapted from (Rowley et al., 2021). When working in the rotating reference frame, the free water pool is saturated at the rate $\boldsymbol{\omega_1 M_Y^A}$, and the bound and dipolar pools' magnetization are saturated at a rate governed by $\boldsymbol{R_{rfB}}$ and $\boldsymbol{\Omega^2 R_{rfB}}$, respectively. Exchange between pools, dictated by **k**, leads to the indirect saturation of the longitudinal magnetization of the free water pool. The magnetization of each pool recovers according to its respective relaxation rates parameters: $\mathbf{R_{1A}}$ ($1/T_{1A}$), $\mathbf{R_{1B}}$ ($1/T_{1B}$), and the relaxation time of the dipolar order $\mathbf{T_{1D}}$.

## S.2 $MT_{sat}$ calculation including turbo-factor and dummy excitations:

For each TR, we model multiple excitations, with spacing ($TR_{exc}$), followed by a time delay TD, and an instantaneous saturation of the bound pool. The total saturation period per TR is added to the true TD to account for relaxation occurring during the RF pulses that we have modelled as an instantaneous event. Additionally, we model dummy excitations following this instantaneous saturation, before taking the pseudo-steady state signal.

For each excitation, the longitudinal magnetization is rotated by flip angle α. For conciseness, we define the change in longitudinal magnetization as $x$, where $x = cos(\alpha)$. Following this, the longitudinal magnetization recovers by a value $y$, where $y = exp\left(\frac{-TR_{exc}}{T_1}\right)$.

The recovery of the longitudinal magnetization at time ***t***, towards the total available magnetization $\boldsymbol{M_0}$ can be modelled as:

$$M_t = M_0(1 - y) + M_{t-1} \cdot y$$

If we take $M_{t-1}$ to be the magnetization immediately following one excitation pulse, the longitudinal magnetization becomes:

$$M_t = M_0(1 - y) + (M_{t-2} \cdot \mathrm{x}) \cdot y$$

$$= M_0 - M_0 \cdot y + M_{t-2} \cdot \mathrm{x} \cdot y$$

This can then be repeated for each excitation/recovery pair, such that the result after three iterations would be:

$$M_t = M_0 - M_0 \cdot y + [M_0 - M_0 \cdot y + (M_0 - M_0 \cdot y + M_{t-6} \cdot \mathrm{x} \cdot y) \cdot \mathrm{x} \cdot y] \cdot \mathrm{x} \cdot y$$

$$= M_{t-6}x^3y^3 + M_0x^2y^2 + M_0xy + M_0 - M_0y - M_0xy^2 - M_0x^2y^3$$

From this, a pattern emerges allowing the longitudinal magnetization after an arbitrary number of repetitions (tf = turbo-factor) to be written as:

$$M_t = M_0 \cdot \sum_{i=1}^{tf} \left[ (x \cdot y)^{i-1} - \frac{(x \cdot \mathrm{y})^i}{x} \right] + M_{t-1} \cdot (x \cdot \mathrm{y})^{tf}$$

It was demonstrated in Munsch et al. (2021), that this can be rewritten as:

$$M_t = M_0 \cdot (1 - \mathrm{y}) \frac{[1 - (x \cdot y)^{tf}]}{(1 - x \cdot y)} + M_{t-1} \cdot (x \cdot \mathrm{y})^{tf}$$

If we separate this into components related to $M_0$ and those dependent on $M_{t-1}$ we can find the magnetization at the end of the turbo-readout ($M_2$) as:

$$M_2 = A_1 + M_1 \cdot B_1$$

$$A_1 = M_0 \cdot (1 - \mathrm{y}) \frac{[1 - (x \cdot y)^{tf}]}{(1 - x \cdot y)} \qquad B_1 = M_1 \cdot (x \cdot \mathrm{y})^{tf}$$

Following the readout, the longitudinal magnetization partially recovers during a delay time ($TD_{app}$ = true TD + saturation time), with the resulting magnetization being equal to:

$M_3 = A_2 + M_2 \cdot B_2$, with:

$$A_2 = M_0 \cdot \left[1 - exp\left(\frac{-TD_{app}}{T_1}\right)\right] \qquad B_2 = exp\left(\frac{-TD_{app}}{T_1}\right)$$

Next, we consider this saturation to be a percentage drop in the steady state signal (δ). The longitudinal magnetization following the train of MT saturation pulses is:

$M_4 = A_3 + M_3 \cdot B_3$, with:

$$A_3 = 0 \qquad B_3 = (1 - \delta)$$

For centric-encoded readouts, dummy excitations are typically played out prior to turning on the ADC. These excitations affect the longitudinal magnetization and are accounted for here. The longitudinal magnetization after *d* dummy excitations is:

$M_5 = A_4 + M_4 \cdot B_4$, with:

$$A_4 = M_0 \cdot (1 - \mathrm{y}) \frac{[1 - (x \cdot y)^d]}{(1 - x \cdot y)} \qquad B_4 = M_4 \cdot (x \cdot \mathrm{y})^d$$

If a pseudo-steady state signal is reached, then $M_5 = M_1$, allowing the system of equations to be solved using:

$$M = \frac{A}{(1 - B)}$$

Where:

$$A = A_4 + A_3B_4 + A_2B_3B_4 + A_1B_2B_3B_4$$

$$B = (1 - B_1B_2B_3B_4);$$

In the absence of dummy excitations (*d* = 0), $A_4 = 0$, and $B_4 = 1$ and the formalism still holds. This can also be looped through for increasing number of dummy excitations to extract the magnetization value that results from each excitation pulse in the turbo-readout.

## S.3 Estimating the impact of the RAGE readout and centric ordering scheme on image intensity:

**Figure S2:** A flow chart depicting the steps used to estimate the impact of the readout on the resulting image intensity. First, the sequence is numerically simulated to obtain magnetization values over the readout. The sampling table is calculated using the image matrix size, and acceleration information such as whether elliptical masking was used, the GRAPPA factor and number of reference lines. Next, the sampling table is filled with the simulated signal values and missing GRAPPA lines are added. The table is replicated in-plane to obtain the desired 3-D matrix size, while ignoring the effects of $T_2$ decay over the readout. The MNI-152 atlas is used as a reference to determine which frequency information most contributes to the image intensity. It is resized and reoriented to the intended image size, and a segmentation is generated using the brain tissue of the atlas, which takes a value of 1 for brain tissue, and 0 elsewhere. This is Fourier transformed to obtain a k-space weighting which is then multiplied with the simulated k-space values. The result is inverse Fourier transformed to obtain an image of the brain tissue that now takes on the simulated signal value. The result is the average value within the initial brain segmentation (voxels with value 1). Images are shown here as axial, with the k-space planes being taken from sagittal plane. The k-space values are log transformed to visualize the high-frequency components.

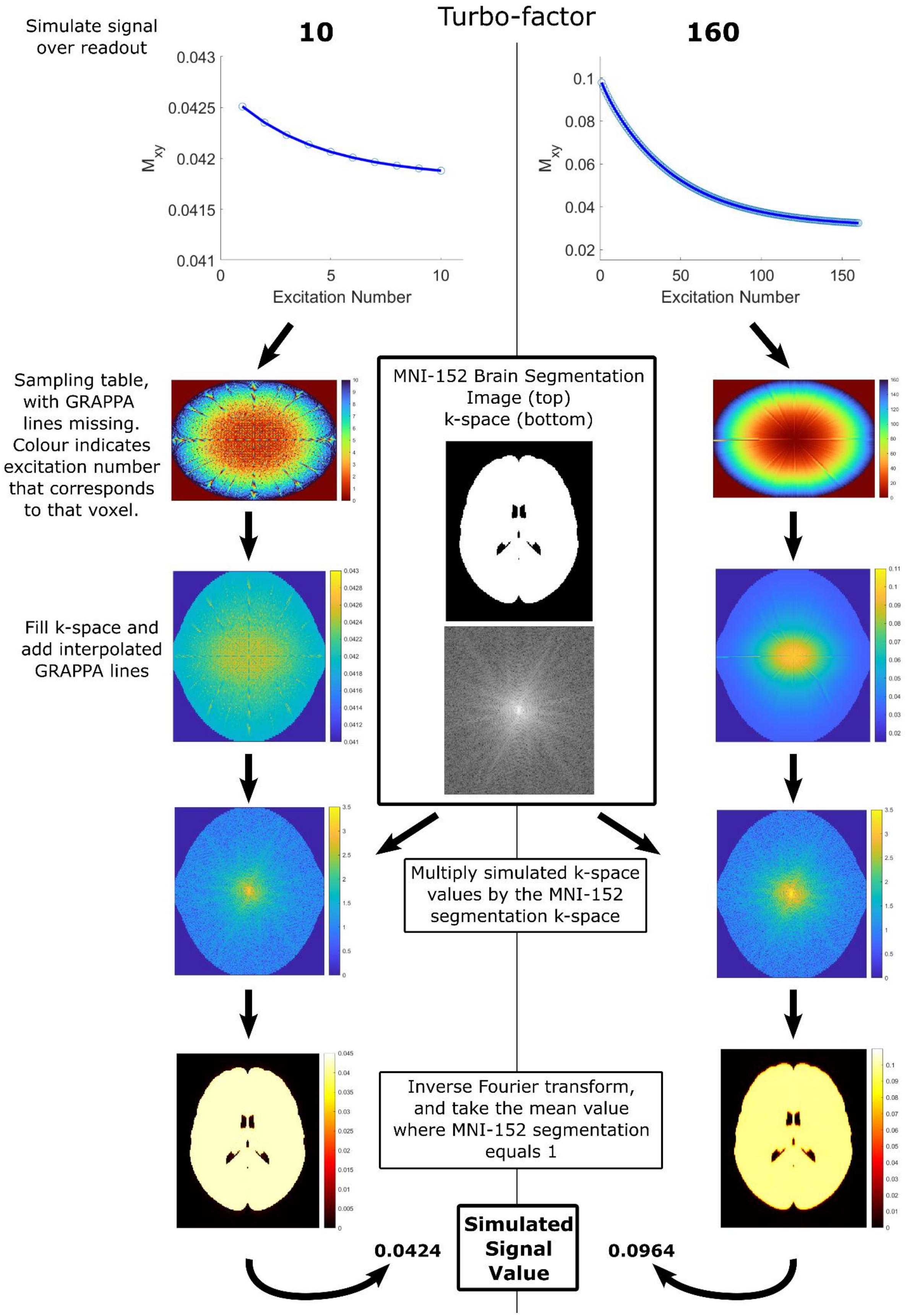
Turbo-factor
10
160
Simulate signal over readout
Mxy
Excitation Number
Sampling table, with GRAPPA lines missing. Colour indicates excitation number that corresponds to that voxel.
Fill k-space and add interpolated GRAPPA lines
MNI-152 Brain Segmentation Image (top) k-space (bottom)
Multiply simulated k-space values by the MNI-152 segmentation k-space
Inverse Fourier transform, and take the mean value where MNI-152 segmentation equals 1
0.0424
Simulated Signal Value
0.0964

## S.4 Noise Calculation:

### S.4.1 MTR:

ihMTR was calculated as:

$$ihMTR = \frac{\left(\frac{Pos+Neg}{2}\right)-Dual}{S_0} = \frac{Pos+Neg}{2S_0} - \frac{Dual}{S_0}$$

Image noise was assumed to have a white gaussian distribution, stemming from thermal system noise, and was fixed to 0.0005 for all images. With a simulated signal value of 0.04, this equates to an SNR level of ~70 of the input image. The standard deviation of the noise (σ) in the resulting ihMTR map will be dependent on the noise and intensity of the input images. This value can be determined analytically if we assume that the errors are random and independent between the collected images by taking the partial derivative with respect to each input image:

$$\sigma_{ihMTR} = \sqrt{\frac{\partial ihMTR}{\partial Pos}^2 \sigma_{Pos}^2 + \frac{\partial ihMTR}{\partial Neg}^2 \sigma_{Neg}^2 + \frac{\partial ihMTR}{\partial Dual}^2 \sigma_{Dual}^2 + \frac{\partial ihMTR}{\partial S_0}^2 \sigma_{S_0}^2}$$

$$= \sqrt{\frac{\sigma_{Pos}^2}{4S_0^2} + \frac{\sigma_{Neg}^2}{4S_0^2} + \frac{\sigma_{Dual}^2}{S_0^2} + \left(\frac{2 \cdot Dual - (Pos + Neg)}{2S_0^2}\right)^2 \sigma_{S_0}^2}$$

SNR was computed as ihMTR/ $\sigma_{ihMTR}$.

### S.4.2 $MT_{sat}$:

With a turbo-factor of 1, the noise of $MT_{sat}$ can be solved analytically. Our approach, with an arbitrary turbo-factor, utilizes a lookup table to solve for $MT_{sat}$. We assumed the same constant noise value as in the ihMTR calculation, and that an MP2RAGE was used to obtain the $M_0$ and $T_1$ values. This means that five separate images are required for calculation of $ihMT_{sat}$ (two from MP2RAGE, and Pos, Neg and Dual MT-w images). MP2RAGE signal was simulated using code available online (https://github.com/JosePMarques/MP2RAGE-related-scripts), with the following parameters: $B_0$ = 3, TR = 5 s, $TR_{FLASH}$ = 6.4e-3 s, TIs = [940e-3, 2830 e-3], slices = 176, and flip degrees = [4, 5]. A 5 x 30,000 noise vector was generated in MATLAB using the *normrnd* function. The value of 30,000 was chosen as it was sufficiently large to consistently provide noise vectors with $\sigma = 0.0005$. Noise was added to the simulated $M_0$ and $T_1$ values to generate 30,000 values of each with noise. The remaining three columns of noise values were used to generate noisy image intensities for the MT-w images. Noisy $MT_{sat}$ was computed for each MT-w image, and then $ihMT_{sat}$ computed as:

$$ihMT_{sat} = MT_{sat,Dual} - \left(\frac{MT_{sat,Pos} + MT_{sat,Neg}}{2}\right)$$

The standard deviation of this $ihMT_{sat}$ vector was calculated to be the resulting noise of the protocol. The SNR was calculated as the $ihMT_{sat}/\ \sigma_{ihMTsat}$.